\documentclass[12pt]{article} 
\usepackage[sectionbib]{natbib}
\usepackage{array,epsfig,fancyheadings,rotating}
\usepackage{lscape, afterpage, color}
\usepackage[]{hyperref}  
\usepackage{sectsty, secdot}
\sectionfont{\fontsize{12}{14pt plus.8pt minus .6pt}\selectfont}
\renewcommand{\theequation}{\thesection\arabic{equation}}
\subsectionfont{\fontsize{12}{14pt plus.8pt minus .6pt}\selectfont}

\usepackage{amsmath}
\usepackage{amssymb}
\usepackage{amsfonts}
\usepackage{multirow}
\usepackage{amsthm}

\newtheorem{theorem}{Theorem}
\newtheorem{lemma}{Lemma}

\newtheorem{proposition}{Proposition}
\theoremstyle{definition}

\newtheorem{example}{Example}
\newtheorem{remark}{Remark}
\allowdisplaybreaks

\begin{document}


\renewcommand{\baselinestretch}{2}

\markright{ \hbox{\footnotesize\rm Statistica Sinica
}\hfill\\[-13pt]
\hbox{\footnotesize\rm
}\hfill }

\markboth{\hfill{\footnotesize\rm G.~Chen, Y.~He, C.~D.~Lin and F.~Sun} \hfill}
{\hfill {\footnotesize\rm GROUPED ORTHOGONAL ARRAYS} \hfill}

\renewcommand{\thefootnote}{}
$\ $\par


\fontsize{12}{14pt plus.8pt minus .6pt}\selectfont \vspace{0.8pc}
\centerline{\large\bf GROUPED ORTHOGONAL ARRAYS AND THEIR}
\vspace{2pt}
\centerline{\large\bf CONSTRUCTION METHODS}
\vspace{.4cm}
\centerline{Guanzhou Chen$^1$, Yuanzhen He$^2$, C.~Devon Lin$^3$ and Fasheng Sun$^4$}
\vspace{.4cm}
\centerline{$^1$\textit{Nankai University}, $^2$\textit{Beijing Normal University},}
\centerline{$^3$\textit{Queen's University and} $^4$\textit{Northeast Normal University}}
 \vspace{.55cm} \fontsize{9}{11.5pt plus.8pt minus.6pt}\selectfont


\begin{quotation}
\noindent {\it Abstract:}
In computer experiments, it has become a standard practice to select the inputs that spread out as uniformly as possible over the design space. The resulting designs are called space-filling designs and they are undoubtedly desirable choices when there is no prior knowledge on how the input variables affect the response and the objective of experiments is global fitting. When there is some prior knowledge on the underlying true function of the system or what statistical models are more appropriate, a natural question is, are there more suitable designs than vanilla space-filling designs? In this article, we provide an answer for the cases where there are no interactions between the factors from disjoint groups of variables. In other words, we consider the design issue when the underlying functional form of the system or the statistical model to be used is additive where each component depends on one group of variables from a set of disjoint groups. For such cases, we recommend using {\em grouped orthogonal arrays.}   Several construction methods are provided and many designs are tabulated for practical use. Compared with existing techniques in the literature, our construction methods can generate many more designs with flexible run sizes and better within-group projection properties for any prime power number of levels.

\vspace{9pt}
\par
\end{quotation}\par

\def\thefigure{\arabic{figure}}
\def\thetable{\arabic{table}}

\renewcommand{\theequation}{\thesection.\arabic{equation}}

\fontsize{12}{14pt plus.8pt minus .6pt}\selectfont

\section{Introduction}
\label{sec:introduction}

Space-filling designs are ubiquitous in the practice of computer experiments and the research community \citep{Santner-Williams-Notz-2018, Gramacy-2020}.
These designs are used for selecting the settings of input variables to
 explore how responses depend on input variables,  by scattering the points in the design region as uniformly as possible.
 In practice,  if there is no preference or knowledge on the choice of statistical models,  a statistically sound way is to collect data from all portions of the design region, that is,  to use space-filling designs to entertain flexible statistical models (also called surrogate models or emulators in computer experiments).  This idea can trace back to as early as \cite{Box-Draper-1959} who introduced a basis for the selection of a response surface design, to the best of our knowledge.  There are rich results on studying construction, optimality and the use of space-filling designs \citep{Lin-Tang-2015, Joseph-2016}.

A new and unexplored territory in designs of computer experiments is, how to choose input settings when there is prior knowledge on the underlying true response surface of the system under study.  Motivated by some recent methodological developments and real applications,  we consider the cases where there are no interactions between the factors from disjoint groups of variables.  In such cases,  the functional form of the true response surface or the preferred surrogate model is additive where each component is a function of one group of variables from a number of disjoint groups.   For example,   in the engine block and head joint sealing experiment discussed in \cite{Joseph-Hung-Sudjianto-2008}, eight factors are selected for experimentation; their analysis revealed that some linear, quadratic and interaction effects of the first, second, and sixth factors are important effects. From the experimental design point of view, we can split the eight factors into two groups, the first, second and sixth factors as one group and the rest as the other group, a design with better projection and space-filling properties for the first group would be more desirable, comparing to space-filling designs without this feature, as it allows more accurate estimation of the main and interaction effects of factors in the first group.

Another application where the proposed designs with a group structure might be useful lies in the use of blocked (or group) additive kernels. Such kernels have been used in  Gaussian processes based on functional analysis of variance decomposition \citep{Muehlenstaedt-etal-2012}, the non-parametric regression \citep{Pan-Zhu-2017}, and Bayesian optimization problems \citep{Gardner-etal-2017}.   An experimental design that takes into account the feature of the kernels would enable more efficient estimation of the parameters in the kernels.

To address these practical needs,  we propose using space-filling designs with a group structure.  In this article, we focus on the projection property onto lower dimensions.  It can be accomplished with an orthogonal array-based Latin hypercube \citep{Owen-1992, Tang-1993} derived from an orthogonal array of high strength.   A good low-dimensional projection property is an essential characteristic of screening designs which are indispensable for more realistic, complex computer experiments as greater numbers of input variables are employed.  In addition, as shown in \cite{Wang-etal-2021}, higher strength guarantees better space-filling properties.  With the group structure, we introduce {\em grouped orthogonal arrays}, so that grouped orthogonal array-based Latin hypercubes can be constructed.  The proposed grouped orthogonal arrays are of strength two but the factors can be divided into disjoint groups with higher strength or minimum aberration (reasons for this property will be given later).  The newly introduced designs extend the concept of designs of variable resolution proposed by \cite{Lin-2012} or variable strength orthogonal arrays by \cite{Raaphorst-Moura-Stevens-2014} in which groups have higher strength than the whole array. \cite{Lin-2012} and \cite{Lekivetz-Lin-2016} provided several constructions for designs of variable resolution but the focus is on two-level designs.
The variable strength orthogonal arrays obtained by \cite{Raaphorst-Moura-Stevens-2014} only have groups of three factors and their run sizes are limited to $s^3$ for a prime power $s$.
\cite{Zhang-Pang-Li-2023} constructed variable strength orthogonal arrays with strength two containing strength greater than two by Galois field and some variable strength orthogonal arrays with strength $l \geq 2$ containing strength greater than $l$ by Fan-construction. In addition to the drawback that the designs constructed have only one group with larger strength, the resulting designs have very restrictive run sizes $s^t$ for a prime power $s$ and an integer $t \geq 4$.

In this paper, we develop several construction methods for the proposed grouped orthogonal arrays.
Compared with existing designs of variable resolution or variable strength orthogonal arrays,  the designs obtained by our methods can have any prime power number of levels and are not restricted to those with strength-three or strength-four groups.
In particular, we first give explicit constructions for grouped orthogonal arrays with strength-three and near strength-three groups, and then explore the constructions of designs with minimum aberration groups in pursuit for better within-group projection properties.
These grouped orthogonal arrays have flexible run sizes as well as various group structures, thereby substantially expanding the designs that could be used for experiments with grouped factors.

The remainder of the paper is organized as follows.
Section \ref{sec:notation} reviews necessary background and introduces the notations to be used.
Section \ref{sec:GOA-strength3} presents construction methods for grouped orthogonal arrays with strength-three and near strength-three groups.
Section \ref{sec:GOA-MA} examines the constructions of grouped orthogonal arrays with minimum aberration groups.
Section \ref{sec:simulation} demonstrates the usefulness of GOAs from two aspects through simulation studies.
Section \ref{sec:conclude} concludes the paper with a discussion on further applications of GOAs and possible directions for future work.
All the proofs,
as well as supplementary design tables
are postponed to the Supplementary Material.

\section{Notations and background}
\label{sec:notation}

In this section, we first review a few important concepts including Galois fields, orthogonal arrays and projective geometries. We refer to the monographs \cite{Dembowski-1968}, \cite{Lidl-Niederreiter-1986} and \cite{Hedayat-Sloane-Stufken-1999} for the in-depth explanations.   Armed with the background knowledge and notations, we then introduce the notion of grouped orthogonal arrays.

Throughout we focus on the case that all factors share the same number of levels and use $s$ to denote the number of levels.
Suppose that $s$ is a prime power and $k$ is a positive integer.
A Galois field of order $s^k$ is a finite field with $s^k$ elements, and is denoted by $GF(s^k) = \{\omega_0, \omega_1, \ldots, \omega_{s^k-1}\}$ with $\omega_0=0$ in this paper.
In particular, if $s$ is a prime number and $k=1$, then the $s$ elements are denoted by $GF(s) = \{0,1,\ldots, s-1\}$.
All the nonzero elements of $GF(s^k)$ can be expressed as powers of an element $\beta \in GF(s^k)$;
that is, $GF(s^k) \setminus \{0\} = \{\beta^0, \beta^1, \ldots, \beta^{s^k-2}\}$.
The element $\beta$ is called a primitive element of $GF(s^k)$.
The primitive polynomial $h(x)$ corresponding to $\beta$ is a polynomial of degree $k$ with coefficients from the subfield $GF(s)$ such that $h(\beta)=0$.
Using addition and multiplication operations of polynomials modulo $h(\beta)$, any nonzero element $\beta^i$ of $GF(s^k)$ for $i=0,1\ldots, s^k-2$ can be written as a linear combination of $1, \beta, \beta^2, \ldots, \beta^{k-1}$ with coefficients $a_{i,j} \in GF(s)$ for $j=1,\ldots, k-1$; that is,
$$
\beta^i = a_{i,0} + a_{i,1} \beta + \cdots + a_{i,k-1} \beta^{k-1}.
$$
Thus the nonzero element $\beta^i$ can also be represented by a vector $(a_{i,0}, a_{i,1}, \\ \ldots, a_{i,k-1})^T \in GF(s)^k$ for $i=0,1\ldots, s^k-2$.
In this paper, we call $(a_{i,0}, a_{i,1}, \ldots, a_{i,k-1})^T$ the vector format and $\beta^i$ the power format for a nonzero element of $GF(s^k)$.

An $N \times n$ matrix with entries from $GF(s)$ is called an orthogonal array of strength $t$, and denoted by OA$(N,n,s,t)$, if any of its $N \times t$ submatrix contains all possible level combinations equally often.
An OA$(N, n, s, t)$ is said to be regular or linear if its $N$ rows form a linear space over $GF(s)$.
A matrix with its rows formed by all vectors in a basis of this linear space is called a generator matrix for the OA$(N, n, s, t)$.
\begin{example}\label{example-0}
The matrix $D$ displayed in the transpose form in \eqref{eqn:linear-OA} is an OA$(8, 4, 2, 3)$.
All rows of $D$ can be generated by linear combinations of
rows of $G$ in \eqref{eqn:linear-OA}.
Thus $G$ is a generator matrix for $D$.

\begin{equation}
\label{eqn:linear-OA}
D = \begin{bmatrix}
0	&	0	&	0	&	0	&	1	&	1	&	1	&	1	\\
0	&	0	&	1	&	1	&	0	&	0	&	1	&	1	\\
0	&	1	&	0	&	1	&	0	&	1	&	0	&	1	\\
0	&	1	&	1	&	0	&	1	&	0	&	0	&	1
\end{bmatrix}^T,
\qquad
  G =
\begin{bmatrix}
1 & 0 & 0 & 1 \\
0 & 1 & 0 & 1 \\
0 & 0 & 1 & 1
\end{bmatrix}.
\end{equation}
\end{example}

A subset of columns of a regular orthogonal array $D$ (or equivalently, its generator matrix) is said to form a defining word if there exist nonzero coefficients from $GF(s)$ such that the linear combinations of these columns is a zero vector.
The length of a defining word is the number of columns involved in it.
Let $A_j$ be the number of defining words of length $j$ in $D$ for any positive integer $j$.
Then the minimum aberration criterion aims to sequentially minimize the entries of the wordlength pattern $(A_1, A_2, A_3, \ldots)$.
In addition, we have $D$ is an OA$(N, m, s, t)$ if and only if $A_j=0$ for $j=1,\ldots, t$.

The columns of a generator matrix for an OA$(s^k, n, s, 2)$ can be seen as points of a projective geometry PG$(k-1,s)$.
These points in $GF(s)^k$ have the property that any two of them are linearly independent over $GF(s)$.
Hence, there are a total of $(s^k-1)/(s-1)$ points in PG$(k-1,s)$.
Given a primitive element $\beta \in GF(s^k)$, all these points are also given by the vector formats of $\beta^i$ for $i=0, \ldots,  (s^k-1)/(s-1)-1$.
A subset of PG$(k-1,s)$ is called a cap if any three points are linearly independent.
Clearly, a matrix using points of a cap as its columns can generate an array with $A_j=0$ for $j=1,2,3$, i.e., an orthogonal array of strength three.

An OA$(N,n,s,t)$ is said to be nonregular if its rows do not form a linear space over $GF(s)$.
The run size of a nonregular orthogonal array can be very flexible as it does not need to be a prime power.
Nonregular orthogonal arrays can be constructed from difference schemes \citep{Hedayat-Sloane-Stufken-1999}.
An $r \times c$ matrix with entries from $GF(s)$ is called a difference scheme if the difference of any two columns contains all elements of $GF(s)$ equally often; we denote this matrix by DS$(r,c,s)$.
We shall revisit this concept in Section \ref{subsec:GOA-recursive}.

In this paper, we study orthogonal arrays where the factors are divided into groups and higher strength, or more generally, minimum aberration, is achieved by factors in the same groups.
We call these arrays the {\em grouped orthogonal arrays} (GOAs).
Aligned with the notation in \cite{Lin-2012}, we use GOA$(N, (m_1,m_2,\ldots, m_g),  (t_1,t_2, \ldots, t_g), s, t_0)$ to denote a GOA of $N$ runs with $g$ groups of $s$-level factors with strength $t_0$ where the $i$th group has $m_i$ factors and is of strength $t_i$, where $t_i \geq t_0$, for $i =1, \ldots, g$.   If all groups have the same size $m$ and are of the same strength $t$, we use GOA$(N, m \times g, t \times g,s, t_0)$.
The two notations may be used in combination; for example, a GOA$(27, (4, 3, 3), 3\times 3, 3,2)$ represents a 27-run design of strength two for $10$ three-level factors, where the 10 factors are divided into three groups each having 4, 3 and 3 factors and of strength three.

\section{GOAs with strength-3 and near strength-3 groups}
\label{sec:GOA-strength3}

This section centers on construction methods for grouped orthogonal arrays with strength-three or near strength-three groups.
As the whole array of these designs is of strength 2, the  subdesign resulting from projecting onto factors within the same group has better space-filling properties.
In addition, such designs allow the main effects to be estimated without being biased by two-factor interactions in the same group.
The merits of these designs will be demonstrated in Section \ref{sec:simulation}.
The construction methods presented in this section are explicit and require little use of computer search. Specifically,
Section \ref{subsec:GOA-s3s4} gives two construction methods based on caps to obtain designs of $s^3$ and $s^4$ runs for any prime power $s$.
Next, a recursive construction method is introduced in Section \ref{subsec:GOA-recursive} to generate designs with larger and flexible run sizes.

\subsection{Designs of $s^3$ and $s^4$ runs}
\label{subsec:GOA-s3s4}
We first present two construction methods for grouped orthogonal arrays of $s^3$ and $s^4$ runs in which the groups have strength 3, where $s$ is a prime power.
The first method uses the following generator matrix $G_0$ which is known as an oval in cap theory,
$$
G_0 = \begin{bmatrix}
1            & 1            & 1            &  \cdots  & 1               & 0\\
 \omega_0    &  \omega_1    &  \omega_2    &  \cdots  &  \omega_{s-1}   & 0\\
 \omega_0^2  &  \omega_1^2  &  \omega_2^2  &  \cdots  &  \omega_{s-1}^2 & 1
\end{bmatrix}.
$$
It can be directly verified that the columns of $G_0$ form a cap in PG$(2, s)$ because any three of them are linearly independent. Thus $G_0$ generates an OA$(s^3, s+1, s, 3)$.
It has been shown that this orthogonal array has the largest number of factors when $s$ is odd, see Corollary 3.9 of \cite{Hedayat-Sloane-Stufken-1999}.
Now let
$$
G_i = \begin{bmatrix}
1            & 1            & 1            &  \cdots  & 1               \\
 \omega_0    &  \omega_1    &  \omega_2    &  \cdots  &  \omega_{s-1}   \\
 \omega_i+\omega_0^2  &  \omega_i+\omega_1^2  &  \omega_i+\omega_2^2  &  \cdots  &  \omega_i+\omega_{s-1}^2
\end{bmatrix}
$$
for $i = 1, \ldots, s-1$. Then we have the following result.
\begin{theorem}
\label{thm:s3-runs}
The generator matrix $G=(G_0, G_1, \ldots, G_{s-1})$ generates a GOA$(s^3, (s+1, s, \ldots, s), 3 \times s, s, 2)$.
\end{theorem}

As an illustration of Theorem \ref{thm:s3-runs}, Example \ref{ex:s3_125} below constructs a five-level GOA with 125 runs and 26 factors that include four groups of 5 factors and one group of 6 factors, and each group is of strength three.

\begin{table}[h]
\caption{The generator matrix $G$ in Example \ref{ex:s3_125}.\label{tab:s3_125}}
\centering
\begin{tabular}{ccccc}
  \hline
$G_0$ & $G_1$ & $G_2$ & $G_3$ & $G_4$\\
  \hline
{111110} & {11111} & {11111} & {11111} & {11111}\\
{012340} & {01234} & {01234} & {01234} & {01234}\\
{014411} & {12002} & {23113} & {34224} & {40330}\\
  \hline
\end{tabular}
\end{table}

\begin{example}
\label{ex:s3_125}
Consider the case $s=5$. According to the result of Theorem \ref{thm:s3-runs}, the generator matrix $G=(G_0, G_1, G_2, G_3, G_4)$ shown in Table \ref{tab:s3_125} generates a GOA$(125, (6, 5, 5, 5, 5), 3\times 5, 3, 2)$.
\end{example}

Next we present a construction method for GOAs of $s^4$ runs.
\cite{Ebert-1985} proved that all the $s^3+s^2+s+1$ points of PG$(3, s)$ can be divided into $s+1$ disjoint caps of size $s^2+1$, which implies the existence of GOA$(s^4, (s^2+1)\times (s+1), 3\times (s+1), s, 2)$ for any prime power $s$.
To present this method, we need to use the power format of points in PG$(3, s)$.
Suppose $\beta$ is a primitive element of $GF(s^4)$.
Then all points of PG$(3, s)$ can be represented by $\beta^0, \beta^1, \ldots, \beta^{s+s^2+s^3}$.
Let $m = s^2 + 1$ and $g = s+1$.
Define $G_0 = (\beta^0, \beta^g, \beta^{2g}, \ldots, \beta^{(m-1)g})$ and $G_i = \beta^i G_0 = (\beta^i, \beta^{g+i}, \beta^{2g+i}, \ldots, \beta^{(m-1)g+i})$ for $i=1, \ldots, g-1$.
The next lemma rephrases Ebert's (\citeyear{Ebert-1985}) result which was originally stated using the language of projective geometry.
\begin{lemma}
\label{lem:s4-runs}
The generator matrix $G=(G_0, G_1, \ldots, G_{g-1})$ generates a GOA$(\\ s^4, (s^2+1)\times (s+1), 3\times (s+1), s, 2)$.
\end{lemma}

We remark that the GOAs obtained by Lemma \ref{lem:s4-runs} have large group sizes, because an OA$(s^4, s^2+1, s, 3)$ attains the largest number of factors among all orthogonal arrays for $s=3$ and among all linear orthogonal arrays for $s>3$, see Section 5.9 of \cite{Hedayat-Sloane-Stufken-1999}.
We conclude this subsection with an illustration of Lemma \ref{lem:s4-runs}.

\begin{example}
\label{ex:s4_81}
Suppose that $\beta$ is a primitive element of $GF(3^4)$, with its primitive polynomial given by $h(x) = x^4 + x + 2$.
According to Lemma \ref{lem:s4-runs}, we define $G_0 = (\beta^0, \beta^4, \beta^8, \ldots, \beta^{36})$ and $G_i = \beta^i G_0$ for $i=1,2,3$.
Then $G=(G_0, G_1, G_2, G_3)$ generates a GOA$(81, 10 \times 4, 3 \times 4, 3, 2)$.
Table \ref{tab:s4_81} displays the columns of $G$ in the vector format.
\end{example}

\begin{table}[h]
\caption{The generator matrix $G$ in Example \ref{ex:s4_81}.\label{tab:s4_81}}
\centering
\begin{tabular}{cccc}
  \hline
$G_0$ & $G_1$ & $G_2$ & $G_3$ \\
  \hline
{1111201121} & {0002220212} & {0010021122} & {0210110202} \\
{0210110202} & {1112011212} & {0022202120} & {0100211220} \\
{0010021122} & {0210110202} & {1112011212} & {0022202120} \\
{0002220212} & {0010021122} & {0210110202} & {1112011212} \\
  \hline
\end{tabular}
\end{table}

\subsection{A recursive construction}
\label{subsec:GOA-recursive}

In this subsection, we introduce a recursive construction method which allows us to obtain GOAs with strength-3 groups for larger run sizes from existing orthogonal arrays and GOAs.
The resulting designs have flexible run sizes as they need not be prime powers.
In addition, the groups of the obtained GOAs can be combined to form larger groups of near strength 3.

In Lemma \ref{lem:DS-OA2} below, we first review a useful result which enables us to generate orthogonal arrays from difference schemes.

\begin{lemma}
\label{lem:DS-OA2}
Suppose that $A$ is a DS$(r,c,s)$ and $B$ is an OA$(N, n, s, 2)$. Then $D = A \oplus B$ is an OA$(Nr, cn, s, 2)$, where $\oplus$ denotes the Kronecker sum in $GF(s)$.
\end{lemma}

Lemma \ref{lem:DS-OA2} is slightly different from the standard result in the literature in which $B \oplus A$ rather than $A \oplus B$ is more often used, see Lemma 6.27 of \cite{Hedayat-Sloane-Stufken-1999}.
Both $A \oplus B$ and  $B \oplus A$ are OA$(Nr, cn, s, 2)$;
as will be seen later, here we use $D=A \oplus B$ so that the group structure of the factors of $D$ becomes immediately clear without re-arranging the columns of $D$.

Given an orthogonal array $D = (d_1, \ldots, d_n)$, we describe its three-column combinatorial orthogonality by the proportion of $(d_i, d_j, d_l)$'s ($1\leq i < j < l \leq n$) with strength 3 and denote it by $p(D)$.
This measure was also used in \cite{He-Lin-Sun-2022}.
Clearly, $D$ has strength 3 if and only if $p(D) = 1$.
The next lemma shows that if the array $B$ in Lemma \ref{lem:DS-OA2} has strength 3, then the resulting design $D$ has a high $p(D)$ value.

\begin{lemma}
\label{lem:DS-OA3}
Suppose that $A$ is a DS$(r,c,s)$ and $B$ is an OA$(N, n, s, 3)$. Then $D = A \oplus B$ is an OA$(Nr, cn, s, 2)$ with
\begin{equation}
\label{eqn:pD}
p(D) \geq 1 - \frac{(c-1)(c-2)}{(cn-1)(cn-2)}.
\end{equation}
In particular, the equality holds as long as $r$ is not a multiple of $s^2$.
\end{lemma}

\begin{remark}
A special case of Lemma \ref{lem:DS-OA3} is related to a result of \cite{He-Lin-Sun-2022} who studied the construction of orthogonal arrays of near strength 3.
By taking $A$ as a DS$(s,s,s)$ and $B$ as an OA$(n_2, m_2, s, 3)$ in Lemma \ref{lem:DS-OA3}, one can obtain a design with the property described by Proposition 4.3 in their paper.
\end{remark}

Lemma \ref{lem:DS-OA3} indicates that when the number of factors $n$ of $B$ is large or when the number of columns $c$ of $A$ is small, the design $D$ will be of near strength 3 since $p(D)$ will approach 1.
In particular, if $c = 1$ or $2$, then $p(D)=1$ and $D$ becomes an OA$(Nr, cn, s, 3)$.
This implies that the design $D$ obtained in Lemma \ref{lem:DS-OA3} is also a GOA with strength-3 groups, as revealed in Proposition \ref{prop:GOA-1}.

\begin{proposition}
\label{prop:GOA-1}
Suppose that $A = (A_1, \ldots, A_{g})$ is a DS$(r,c,s)$, where each $A_i$ contains one or two columns for $i=1, \ldots, g$, and $B$ is an OA$(N, n, s, 3)$. Let $D_i = A_i \oplus B$.
Then $D = A \oplus B = (D_1, \ldots, D_g)$ is a GOA with $g$ strength-3 groups each having $n$ or $2n$ columns.
\end{proposition}

Example \ref{ex:DS-OA3-1} below illustrates Proposition \ref{prop:GOA-1} by constructing a GOA$(2s^5,\\ (2s^2 + 2) \times s, 3 \times s, s, 2)$ for any prime power $s$.

\begin{example}
\label{ex:DS-OA3-1}
Take $A=(A_1, A_2, A_3)$ as the DS$(6, 6, 3)$ displayed in Table 6.37 of \cite{Hedayat-Sloane-Stufken-1999}, where each $A_i$ consists of 2 columns for $i=1,2,3$, and take $B$ as an OA$(81, 10, 3, 3)$. Then $D = (D_1, D_2, D_3)$, where $D_i = A_i \oplus B$, is a GOA$(486, 20 \times 3, 3 \times 3, 3, 2)$.
More generally,
from the DS$(2s, 2s, s)$ in Theorem 6.33 of \cite{Hedayat-Sloane-Stufken-1999} and an OA$(s^4, s^2+1, s, 3)$, a GOA$(2s^5, (2s^2 + 2) \times s, 3 \times s, s, 2)$ can be obtained for any prime power $s$.
\end{example}

If larger group sizes are needed, we can partition the columns of $A$ in Lemma \ref{lem:DS-OA3} into larger blocks to obtain larger groups with the near strength 3 property in the resulting design $D$, as illustrated in Example \ref{ex:DS-OA3-2}.

\begin{example}
\label{ex:DS-OA3-2}
Consider the same $A$ and $B$ as in Example \ref{ex:DS-OA3-1}.
If we partition $A$ as $A = (\tilde{A}_1, \tilde{A}_2)$ where $\tilde{A}_i$ has 3 columns for $i=1,2$,
then the design $D = (\tilde{D}_1, \tilde{D}_2)$ is a GOA$(162, 12 \times 2, 2 \times 2,3, 2)$ where $\tilde{D}_i = \tilde{A}_i \oplus B$ has $p(\tilde{D}_i) = 1 - (2\times 1)/(11 \times 10) = 98.2\%$ for $i=1,2$.
\end{example}

By taking $B$ in Lemma \ref{lem:DS-OA3} as GOAs, we obtain a more general recursive construction for GOAs with strength-3 and near strength-3 groups.

\begin{theorem}
\label{thm:GOA-2}
Suppose $A$ is a DS$(r,c,s)$ and $B = (B_1, \ldots, B_g)$ is a GOA$(N, \\ (m_1, \ldots, m_g), 3 \times g, s, 2)$. Let $D_i = A \oplus B_i$. Then $D = (D_1, \ldots, D_g)$ is a GOA$(Nr, (cm_1, \ldots, cm_g), 2 \times g, s, 2)$ where the proportion of strength-3 groups of three columns, $p(D_i)$, of the $i$th group $D_i$ satisfies
$$
p(D_i) \geq 1 - \frac{(c-1)(c-2)}{(cm_i-1)(cm_i-2)}.
$$
Furthermore, as indicated by Proposition \ref{prop:GOA-1}, $D_i$ is also a GOA with strength-three groups in which each group has $m_i$ or $2m_i$ columns, for $i=1,\ldots,g$.
\end{theorem}

Theorem \ref{thm:GOA-2} is a direct consequence of Lemma \ref{lem:DS-OA3} and Proposition \ref{prop:GOA-1} and thus we omit its proof.
It can be shown that $D_i$'s in Theorem \ref{thm:GOA-2} have strength 3 for $s=2$. This special case was previously investigated by Proposition 1 of \cite{Lin-2012}.
We illustrate Theorem \ref{thm:GOA-2} with an example.

\begin{example}
Through a computer search (details are provided in Section S3 of the Supplementary Material), we found a GOA$(54, (5, 5, 4, 4, 4), 3\times 5, 3, 2)$ from the OA$(54,25,3,2)$ documented on page 59 of \cite{Hedayat-Sloane-Stufken-1999}. Applying Theorem \ref{thm:GOA-2} to a DS$(6,6,3)$ and this GOA, we obtain a GOA$(324, (30, 30, 24, 24, 24), 2 \times 5, 3, 2)$ where the groups $D_i$'s satisfy $p(D_i) = 97.5\%$ for $i=1,2$ and $p(D_i) = 96.0\%$ for $i=3,4,5$.
On the other hand, by Proposition \ref{prop:GOA-1}, it is also a GOA$(324, (10\times 6, 8 \times 9), 3 \times 15, 3, 2)$.
\end{example}

\section{GOAs with minimum aberration groups}
\label{sec:GOA-MA}

As a refinement of the criterion of strength, the minimum aberration criterion can be used to select an orthogonal array with better projection properties \citep{Chen-1998, Tang-2001, Wang-etal-2021}.
The aim of this section is to construct regular grouped orthogonal arrays in which the groups have minimum aberration among all regular orthogonal arrays.
Such designs are useful for situations where GOAs with strength-3 and near strength-3 groups do not exist, and where there are many GOAs with strength-3 groups and it is desirable to use one with better within-group projection properties.
Two construction methods will be presented in Sections \ref{subsec:consecutive-powers} and \ref{subsec:algorithm}, respectively.


We consider regular designs throughout this section.
Suppose that $\beta$ is a primitive element of $GF(s^k)$.
As mentioned in Section \ref{sec:notation}, the $(s^k-1)/(s-1)$ columns of PG$(k-1, s)$ can be expressed as
\begin{equation}
\label{eqn:powers-order}
\beta^0, \beta^1, \beta^2, \ldots, \beta^{(s^{k}-1)/(s-1)-1}.
\end{equation}
The following lemma plays an instrumental role in this section.

\begin{lemma}
\label{lem:isomorphism}
For any positive integers $m$ and $j$ and any sequence of increasing integers $ i_1 < \cdots < i_m$, the two designs generated by $(\beta^{i_1}, \ldots, \beta^{i_m})$ and $(\beta^{i_1+j}, \ldots, \beta^{i_m+j})$ have the same wordlength pattern.
\end{lemma}

\subsection{Forming groups by selecting consecutive powers of $\beta$}
\label{subsec:consecutive-powers}

We form groups of a GOA by taking $m$ consecutive elements in \eqref{eqn:powers-order}; that is, the first group is given by $(\beta^0, \beta^1, \ldots, \beta^{m-1})$, the second group is given by $(\beta^m, \beta^{m+1}, \ldots, \beta^{2m-1})$ and so on. According to Lemma \ref{lem:isomorphism}, all groups have the same wordlength pattern.
The structure of each group is established by Theorem \ref{thm:consecutive-powers}.

\begin{theorem}
\label{thm:consecutive-powers}
For a primitive element $\beta$ of $GF(s^k)$,
let $D_0=(d_1, d_2, \ldots, d_m)$ be generated by any $m$ consecutive columns in \eqref{eqn:powers-order}.
Then we have that
(i) if $m \leq k$, then $D_0$ is an OA$(s^k, m, s, m)$; and
(ii) if $m > k$, then $D_0$ is a fractional factorial design, with the defining relation given by
$$
I = d_1^{b_0} d_2^{b_1}\cdots  d_{k+1}^{b_k} =
d_2^{b_0} d_3^{b_1}\cdots d_{k+2}^{b_k}
= \cdots\cdots
= d_{m-k}^{b_0} d_{m-k+1}^{b_1}\cdots d_{m}^{b_k},
$$
where, for example, $I = d_1^{b_0} d_2^{b_1}\cdots  d_{k+1}^{b_k}$ represents the defining word $b_0 d_1 + b_1 d_2 + \cdots b_k d_{k+1} = 0$, and $b_0, b_1, \ldots, b_k \in GF(s)$ are the coefficients of the primitive polynomial $h(x) =  b_k x^k + \cdots + b_1 x + b_0$ corresponding to $\beta$.
\end{theorem}

When $m \leq k$, part (i) of Theorem \ref{thm:consecutive-powers} provides a construction method of GOA$(s^k, m \times \lfloor (s^k-1)/(ms-m)\rfloor, m \times \lfloor (s^k-1)/(ms-m)\rfloor, s, 2)$, where $\lfloor \cdot \rfloor$ denotes the floor function, such that each group contains $s^k/s^m$ replicate(s) of the full factorial.

Now we focus on the case $m > k$. Part (ii) of Theorem \ref{thm:consecutive-powers} shows that the primitive polynomial $h(x) = b_k x^k + \cdots + b_1 x + b_0$ of $\beta$ determines the defining words of $D_0$.
The next result describes $h(x)$ such that $D_0$ has minimum aberration for $m=k+1$ and $m=k+2$.
Some notations are needed to capture certain characteristics of $h(x)$.
Define $b_{-1}=b_{k+1}=0$ and let $f_i=|\{j: b_{j-1} \neq 0, b_j/b_{j-1} = \omega_i, j=0, \ldots, k+1\}|$ for $i=0,1,\ldots, s-1$, $f_{s}=|\{j: b_{j-1} = 0, b_j \neq 0, j=0, \ldots, k+1\}|$ and $f_*=|\{j: b_{j-1} = b_j = 0, j=0, \ldots, k+1\}|$, where $|A|$ denotes the cardinality of a set $A$.

\begin{proposition}
\label{prop:MA-pp}
We have the following results on $D_0$ and $h(x) = b_k x^k + \cdots + b_1 x + b_0$.
\begin{itemize}
\item[(i)] Suppose $m=k+1$.
Then a minimum aberration design $D_0$ maximizes the number of nonzero elements among $b_0, b_1, \ldots, b_{k-1}$. In particular, if $b_0, b_1, \ldots, b_{k-1}$ are all nonzero, then $D_0$ has minimum aberration among all regular OA$(s^k, m, s, 2)$'s.
\item[(ii)] Suppose $m=k+2$. Let $f_{(0)} \leq f_{(1)} \leq \cdots \leq f_{(s)}$ be the ordered values of $f_0, f_1, \ldots, f_{s}$.
Then a minimum aberration design $D_0$ minimizes $f_*+f_{(s)}, \ldots, f_*+f_{(0)}$ sequentially. In particular, if $f_*=0$ and $f_0, f_1, \ldots, f_{s}$ differ by at most 1, then $D_0$ has minimum aberration among all regular OA$(s^k, m, s, 2)$'s.
\end{itemize}
\end{proposition}

We illustrate the results of Proposition \ref{prop:MA-pp} through an example.

\begin{example}
Table C in Chapter 10 of \cite{Lidl-Niederreiter-1986} gives all 22 primitive polynomials for $GF(3^5)$. Among them, we have that
\begin{itemize}
\item[(i)] 4 primitive polynomials satisfy the condition in part (i) of Proposition \ref{prop:MA-pp} and thus lead to GOA$(243, 6 \times 20, 5 \times 20, 3, 2)$'s with minimum aberration groups. For instance, one such polynomial is $h(x) = x^5 + x^4 + x^3 + x^2 + 2x + 1$.
\item[(ii)] 6 primitive polynomials satisfy the conditions in part (ii) of Proposition \ref{prop:MA-pp} and thus lead to GOA$(243, 7 \times 17, 4 \times 17, 3, 2)$'s with minimum aberration groups. For instance, one such polynomial is $h(x) = x^5 + x^3 + 2x^2 + 2x + 1$.
\end{itemize}
\end{example}

Next, we conduct a computer search for the primitive polynomials $h(x)$ that minimize the aberration of $D_0$ in Theorem \ref{thm:consecutive-powers}.
Our search is complete for $s=2,3,5,7$ and all $k$'s such that  the run size $s^k$ is less than 1000 and $k < m \leq k+4$, because all primitive polynomials of $GF(s^k)$ have been checked.
The search results are provided in Tables S1 and S2 in the Supplementary Material, which give the truncated wordlength pattern $(A_3, A_4, A_5, A_6)$ of the groups of all the obtained GOA$(s^k, m \times \lfloor (s^k-1)/(ms-m)\rfloor, t \times \lfloor (s^k-1)/(ms-m)\rfloor, s, 2)$'s ($m \geq t \geq 2$).
In addition, a GOA is marked with an asterisk if its groups have minimum aberration among all regular designs,
which is verified either by Proposition \ref{prop:MA-pp} or by comparison with existing minimum aberration designs in the literature \citep{Chen-Wu-1991, Xu-2005}.
For each GOA in these tables, we give the coefficients $(b_k, b_{k-1}, \ldots, b_1, b_0)$ of one possible $h(x)$ which can generate the design.

\subsection{Forming groups by an algorithm}
\label{subsec:algorithm}

The groups of a GOA obtained in Section \ref{subsec:consecutive-powers} may not have minimum aberration among all regular designs.
For example, the groups of the GOA$(32, 6 \times 5, 4 \times 5, 2, 2)$ presented in Table S1 have a defining word of length 5, while the minimum aberration regular OA$(32, 6, 2, 5)$ only has one defining word of length 6.
In this subsection, we use an algorithm to obtain GOAs whose groups are guaranteed to have minimum aberration among all regular designs.

Given a primitive polynomial, we can express the columns of a generator matrix as powers of a primitive element.
The idea of our algorithm is to use an existing regular minimum aberration design as the first group, then transform the columns of its generator matrix into power formats and apply Lemma \ref{lem:isomorphism} to find the remaining groups.
However, it should be noted that the set of power formats for a minimum aberration design is not unique.
On one hand, there exist many primitive polynomials for a Galois field.
On the other hand, a minimum aberration design can be generated from various generator matrices.
To see this, consider the following two generator matrices for a minimum aberration OA$(16, 5, 2, 4)$:
$$
G_1 = \begin{bmatrix}
1	&	0	&	0	&	0	&	1	\\
0	&	1	&	0	&	0	&	1	\\
0	&	0	&	1	&	0	&	1	\\
0	&	0	&	0	&	1	&	1	\\
\end{bmatrix}
\quad \mbox{and} \quad
G_2 = \begin{bmatrix}
1	&	1	&	1	&	0	&	1	\\
0	&	0	&	0	&	1	&	1	\\
1	&	0	&	1	&	1	&	1	\\
1	&	1	&	0	&	0	&	0	\\
\end{bmatrix}.
$$
If we label the five columns by $r_1, r_2, r_3, r_4$ and $r_5$, then $r_5$ can be seen as generated from the other four independent columns according to $r_5=r_1+r_2+r_3+r_4$.
Hence the generator matrices $G_1$ and $G_2$ generate two designs with the identical wordlength pattern, but they are distinct because two different sets of independent columns $(r_1, r_2, r_3, r_4)$ are used.
Clearly, the power formats for the columns of $G_1$ and $G_2$ are different in general.
Therefore, given a $k\times m$ generator matrix $G$ for a minimum aberration design, other \textit{equivalent} generator matrices $HG$ should also be taken into account, where $H$ is a $k\times k$ non-singular matrix over $GF(s)$.
Based on the discussion above, we describe our full algorithm as follows.
\begin{enumerate}
\item Obtain a generator matrix $G$ for a regular OA$(s^k, m, s, t)$ with minimum aberration.
\item Choose a primitive polynomial $h(x)$ for $GF(s^k)$ and generate a random non-singular $k \times k$ matrix $H$ over $GF(s)$.
Write the columns of $G_0 \leftarrow HG$ as powers of a primitive element $\beta$ associated with $h(x)$, say $(\beta^{i_1}, \beta^{i_2}, \cdots, \beta^{i_m})$.
Set $v \leftarrow (s^k-1)/(s-1)$ and $\mathcal{G} \leftarrow (\beta^{i_1 \bmod v}, \beta^{i_2\bmod v}, \cdots, \beta^{i_m\bmod v})$, where $\bmod$ is the modulo operator.
\item Set $g \leftarrow 1$. For $j=1,\ldots, v-1$:
\begin{enumerate}
\item Let $G_j \leftarrow (\beta^{(i_1+j)\bmod v}, \beta^{(i_2+j)\bmod v}, \cdots, \beta^{(i_m+j)\bmod v})$.
\item If $\mathcal{G} \cap G_j = \emptyset$, then let $\mathcal{G} \leftarrow \mathcal{G} \cup G_j$ and $g \leftarrow g + 1$.
\end{enumerate}
\item Repeat Steps 2 and 3 for a large number $R$ of times and output the largest $g$ as well as the corresponding GOA$(s^k, m \times g, t \times g, s, 2)$.
\end{enumerate}

We set $R=100,000$ and then apply our algorithm to regular minimum aberration OA$(2^k, m, 2, t)$'s for $k=4,5,6$ and $m < 2^{k-1}$ \citep{Chen-Sun-Wu-1993}, OA$(128, m, 2, t)$'s for $m \leq 20$ \citep{Block-Mee-2005} and OA$(3^k, m, 3, t)$'s for $k=4,5$ and $m \leq 20$ \citep{Xu-2005}.
The resulting GOA$(s^k, m \times g, t \times g, s, 2)$'s are presented in Tables S3, S4, S5 and S6 in the Supplementary Material.
Some cases are omitted, either because better designs have been reported in Tables S1 and S2 or because the algorithm cannot find designs with $g \geq 2$.
In addition, a GOA is marked by a dagger if the maximum number of groups is attained, i.e., $g = \lfloor (s^k-1)/(ms-m)\rfloor$.

We conclude this section with some remarks on the two methods of generating GOAs with minimum aberration groups.
Compared with the method of selecting consecutive powers in Section \ref{subsec:consecutive-powers}, the algorithm proposed in Section \ref{subsec:algorithm} can generate GOAs whose groups have less aberration.
On the other hand, the algorithm relies on the existing regular minimum aberration designs and may not yield the maximum number of groups due to the computational burden.
For example, in the algorithmic search we can only find a GOA$(243, 6 \times 18, 5 \times 18, 3, 2)$, while Table S1 gives a GOA$(243, 6 \times 20, 5 \times 20, 3, 2)$ with the same wordlength pattern by choosing an appropriate primitive polynomial.

\section{Simulation and comparison results}
\label{sec:simulation}

In this section, we demonstrate the usefulness of GOAs from two different aspects through simulation studies.
Specifically, Section \ref{subsec:bias} shows GOAs lead to more accurate estimation for main effects when data are generated via an additive regression model, while Section \ref{subsec:prediction} illustrates the advantage of GOAs in providing better predictions in computer experiments.

\subsection{Main effects estimation}
\label{subsec:bias}

In this subsection, we will demonstrate that the GOAs are useful for situations where there are no interactions between factors from disjoint groups of variables.
Suppose that a total of $n=\sum_{k=1}^g m_k$ factors are divided into $g$ disjoint groups where the factors in the $k$th group are denoted by $x_1^{(k)}, \ldots, x_{m_k}^{(k)}$ for $k=1,\ldots,g$. We assume the response $y$ and the factors $x_j^{(k)}$'s are linked through an additive regression model,
\begin{equation}
\label{eqn:true_model_bias}
y = \beta_0 + \sum_{k=1}^g f_k(x_1^{(k)}, \ldots, x_{m_k}^{(k)}) + \epsilon,
\end{equation}
where $\beta_0$ is an intercept, $\epsilon$ is a random error and $f_k$ is a function including linear and quadratic main effects as well as associated two-factor interactions of factors belonging to the $k$th group, that is,
\begin{multline}
\label{eqn:f_k}
f_k(x_1^{(k)}, \ldots, x_{m_k}^{(k)}) = \sum_{j=1}^{m_k}\left(\beta_{j,l}^{(k)} x_{j,l}^{(k)} +\beta_{j,q}^{(k)} x_{j,q}^{(k)} \right) +
\sum_{j_1<j_2} \left(
\beta_{j_1j_2,ll}^{(k)} x_{j_1,l}^{(k)} x_{j_2,l}^{(k)} \right.\\
\left.+\beta_{j_1j_2,lq}^{(k)} x_{j_1,l}^{(k)} x_{j_2,q}^{(k)}+
\beta_{j_1j_2,ql}^{(k)} x_{j_1,q}^{(k)} x_{j_2,l}^{(k)}+
\beta_{j_1j_2,qq}^{(k)} x_{j_1,q}^{(k)} x_{j_2,q}^{(k)}
\right),
\end{multline}
where we assume that all factors $x_j^{(k)}$'s have been scaled to $[0, 2]$ and $x_{j,l}^{(k)} = \sqrt{6}(x_j^{(k)}-1)/2$ and $x_{j,q}^{(k)} = \sqrt{2}\{ 3(x_j^{(k)}-1)^2 - 2\}/2$.

Suppose that our primary interest lies in estimating the linear and quadratic main effects $\beta_{j,l}^{(k)}$ and $\beta_{j,q}^{(k)}$ despite the non-negligible two-factor interactions $\beta_{j_1j_2,ll}^{(k)}$, $\beta_{j_1j_2,lq}^{(k)}$, $\beta_{j_1j_2,ql}^{(k)}$ and $\beta_{j_1j_2,qq}^{(k)}$.
In our simulation study, we first generate these factorial effects and the intercept $\beta_0$ from normal distributions:
$\beta_0, \beta_{j,l}^{(k)}, \beta_{j,q}^{(k)} \sim \mathcal{N}(0, 10^2)$ and $\beta_{j_1j_2,ll}^{(k)},\beta_{j_1j_2,lq}^{(k)}, \beta_{j_1j_2,ql}^{(k)},\beta_{j_1j_2,qq}^{(k)}  \sim \mathcal{N}(0, \sigma^2)$.
Then given a design $D$ of $N$ runs for the $n$ factors, let $X$ be the model matrix corresponding to intercept and the main effects and $Y$ be the vector of $N$ independent responses generated from model \eqref{eqn:true_model_bias}, where, for simplicity, the random errors $\epsilon$ are independently generated from $\mathcal{N}(0, 1)$.
By ignoring the interactions, one can estimate the vector $\beta$ of main effects (including all $\beta_{j,l}^{(k)}$'s and all $\beta_{j,q}^{(k)}$'s) by removing the first entry of $(X^T X)^{-1} X^T Y$.
Denote this estimate by $\hat{\beta}$.
Then the error in $\hat{\beta}$ can be evaluated by the mean squared error $e(\hat{\beta}) = \{\|\hat{\beta} - \beta\|_2^2/(2n)\}^{1/2}$ and a design is desirable if $e(\hat{\beta})$ is minimized on average.

Two specific cases are investigated to illustrate the performance of GOAs.
In case (i), we set $N=27$, $g=3$, $m_1=4$ and $m_2=m_3=3$, for which case we take the design $D$ as a GOA$(27, (4, 3, 3), 3\times 3, 3,2)$ generated by Theorem \ref{thm:s3-runs}. In case (ii), we set $N=81$, $g=3$ and $m_1=m_2=m_3=4$, for which case we take the design $D$ as a GOA$(81, 4 \times 3, 3\times 3, 3,2)$ generated by Lemma \ref{lem:DS-OA3}.
In both cases, we also take $D$ as obtained by randomly permuting columns of the regular minimum aberration designs \citep{Xu-2005} for comparison.
Then the procedure described in the last paragraph is repeated 1000 times, and the average of the 1000 $e(\hat{\beta})$ values is displayed in Table \ref{tab:bias-rmse} for various settings of $\sigma^2$.

\begin{table}[h]
\caption{The average $e(\hat{\beta})$ values obtained from 1000 replications, where the numbers in the parentheses are the corresponding standard errors. \label{tab:bias-rmse}}
\centering
\resizebox{\columnwidth}{!}{\begin{tabular}{cccc}
\hline
 & $\sigma=1$ & $\sigma=5$ & $\sigma=10$ \\
\hline
GOA$(27, (4, 3, 3), 3\times 3, 3,2)$ & 1.223 (0.007) & 6.091 (0.033) & 12.009 (0.065)\\
MA OA$(27,10,3,2)$ & 1.283 (0.008) & 6.323 (0.038) & 12.623 (0.038)\\
  \hline
GOA$(81, 4 \times 3, 3\times 3, 3,2)$ & 0.110 (0.001) & 0.110 (0.001) & 0.109 (0.001)\\
MA OA$(81,12,3,2)$ & 0.495 (0.001) & 2.417 (0.033) & 4.845 (0.061)\\
\hline
\end{tabular}}
\end{table}

From Table \ref{tab:bias-rmse}, it is evident that compared to minimum aberration designs, the GOAs result in smaller $e(\hat{\beta})$ values and hence more accurate main effects estimation.
It is interesting to note that the magnitude of two-factor interactions does not affect the average $e(\hat{\beta})$ values for the GOA$(81, 4 \times 3, 3\times 3, 3,2)$.
This is implied in the proof of Lemma \ref{lem:DS-OA3}, from which one can conclude that the main effects are in fact clear of all two-factor interactions in model \eqref{eqn:true_model_bias}.

On the other hand, for the GOA$(27,(4,3,3), 3\times 3, 3, 2)$ under study, a main effect might still be biased by a two-factor interaction from another group.
Nevertheless, by making the main effects clear of two-factor interactions from the same group, a GOA tends to result in less bias in main effects estimation, as can be seen from the results of Table \ref{tab:bias-rmse}.
This is the intuition behind the theoretical arguments of \cite{Lin-2012}.
As these arguments will be more tedious and complicated than those for two-level designs, we choose to leave it for future work.


\begin{remark}
\label{remark:group-effect-sparsity}
The GOAs are also useful for the scenario of ``group-effect sparsity'', which means that among all groups of factors, perhaps only a few groups are active and really affect the response. More simulation studies are presented in the Supplementary Material to show the advantage of GOAs in this scenario.
\end{remark}

\subsection{Predictions in computer experiments}
\label{subsec:prediction}

In this subsection, we show that GOAs can be used to provide more accurate predictions in computer experiments.
We adopt the same notation as in Section \ref{subsec:bias} and assume the true underlying model can be written as
\begin{equation}
\label{eqn:true_model_CE}
y = \beta_0 + \sum_{k=1}^g f_k(x_1^{(k)}, \ldots, x_{m_k}^{(k)}),
\end{equation}
where $f_k$ takes the same form as in \eqref{eqn:f_k}.
The only difference between models \eqref{eqn:true_model_bias} and \eqref{eqn:true_model_CE} is that a random error
is not included in \eqref{eqn:true_model_CE} as the model is assumed to be deterministic.

Given model \eqref{eqn:true_model_CE} and a design, we can generate a training dataset in the same way as described in the last subsection.
We fit the training dataset with a universal kriging model with fixed linear and quadratic main effects and the Gaussian correlation function \citep{Gramacy-2020}.
Next, a random Latin hypercube design of $N_t=1000$ runs is used as a test dataset, and the fitted model is used to predict the corresponding outputs.
The R package \texttt{DiceKriging} \citep{Roustant-Ginsbourger-Deville-2012} is used to implement the model fitting and prediction procedure described above.
Let $x_i$ be the $i$th point of the test dataset, $y(x_i)$ be the true response from \eqref{eqn:true_model_CE}, and $\hat{y}(x_i)$ be the prediction obtained from the fitted model.
Then the prediction error can be evaluated by the root mean squared error $\mbox{RMSE}=\big[\sum_{i=1}^{N_t}\{y(x_i) - \hat{y}(x_i)\}^2/N_t\big]^{1/2}$.

Again, we examine the performances of GOA$(27, (4, 3, 3), 3\times 3, 3,2)$ and GOA$(81, 4 \times 3, 3\times 3, 3,2)$ by comparing them with other types of designs.
In addition to the minimum aberration designs (MAOAs) considered in Section \ref{subsec:bias}, we also study Latin hypercube designs obtained by randomly expanding the levels of GOAs and MAOAs, the maximum projection Latin hypercube designs \citep{Joseph-Gul-Ba-2015} generated by the \texttt{MaxProLHD()} function in R package \texttt{MaxPro} and the maximin-distance Latin hypercube designs \citep{Ba-Myers-Brenneman-2015} generated by the \texttt{maximinSLHD()} function in R package \texttt{SLHD}.
For each of these designs, we repeat the procedure described in the last paragraph 1000 times, and summarize the averages of the 1000 RMSE values in Table \ref{tab:pred-rmse}.

The simulation results show that GOAs outperform other designs.
It can also be observed that the GOAs and MAOAs tend to have lower RMSE values than Latin hypercube designs.
This is explainable because the three-level designs allow more accurate estimation of the linear and quadratic main effects in the fitted model, which will help make better predictions at unexplored locations.
Nonetheless, the Latin hypercube designs are more desirable when the underlying functions are highly complex and many levels are needed to fully explore the response surface.
In Table \ref{tab:pred-rmse}, it can be seen that the GOA-based Latin hypercube designs have superior performance among different types of Latin hypercube designs.

\begin{table}
\caption{The average RMSE values obtained from 1000 replications, where the numbers in the parentheses are the corresponding standard errors. \label{tab:pred-rmse}}
\centering
\resizebox{\columnwidth}{!}{\begin{tabular}{cccc}
\hline
 & $\sigma=1$ & $\sigma=5$ & $\sigma=10$ \\
\hline
GOA$(27, (4, 3, 3), 3\times 3, 3,2)$ & 3.860 (0.015) & 19.238 (0.072) & 38.053 (0.149)\\
MAOA & 3.936 (0.015) & 19.542 (0.075) & 38.664 (0.157)\\
GOA-based LHD & 5.374 (0.033) & 26.455 (0.160) & 53.313 (0.317)\\
MAOA-based LHD & 5.432 (0.034) & 26.613 (0.159) & 53.510 (0.319)\\
MaxPro LHD & 5.695 (0.045) & 28.601 (0.224) & 57.278 (0.453)\\
maximin LHD & 7.872 (0.112) & 40.218 (0.642) &  70.870 (1.280)\\
  \hline
GOA$(81, 4 \times 3, 3\times 3, 3,2)$ & 3.882 (0.010) & 19.434 (0.055) & 38.879 (0.108)\\
MAOA & 4.020 (0.011) & 20.098 (0.059) & 40.259 (0.117)\\
GOA-based LHD & 4.627 (0.016) & 23.031 (0.082) & 46.301 (0.161) \\
MAOA-based LHD & 4.688 (0.016) & 23.357 (0.082) & 46.721 (0.167)\\
MaxPro LHD & 4.644 (0.015) & 23.356 (0.081) & 46.692 (0.159)\\
maximin LHD & 5.492 (0.043) & 25.467 (0.200) & 54.232 (0.404)\\
\hline
\end{tabular}}
\end{table}

\section{Concluding remarks}
\label{sec:conclude}

In this paper, we introduce grouped orthogonal arrays for experiments where the factors under investigation are divided into groups and better projection properties are desired for factors belonging to the same group.
We present explicit constructions for GOAs with strength-3 groups for $s^3$ and $s^4$ runs,
as well as a recursive construction method which enables us to obtain large GOAs with strength-3 and near strength-3 groups.
Then we judiciously select powers of a primitive element to obtain regular GOAs with minimum aberration groups.
 
\subsection{Further applications of GOAs}

The GOAs share some interesting connections with experimental designs used for group screening \citep{Lewis-Dean-2001,Moon-Dean-Santner-2012,Draguljic-etal-2014}.
The group screening method identifies active factors by first dividing all potential factors into groups and determining the active groups, then examining the factors belonging to those active groups.
It is interesting to note that GOAs can be applied to study such problems, especially if prior knowledge indicates that factors in the same group are more likely to interact with each other.
For example, \cite{Sexton-Lewis-Please-2001} documented an experiment which was aimed at improving the performance of hydraulic gear pumps.
A pump is usually made up of several components where each component consists of some geometrical features that could impact the response.
A GOA can be naturally employed by treating each feature as a factor and each component as a group.
For another example, in industrial processes involving a series of stages, the quality of the final product might be affected by several factors at each stage \citep{Tyssedal-Kulahci-2015}.
Then factors of the same stage can be viewed as coming from the same group.
Once the active groups have been identified, the projection properties of GOAs provide benefits for studying the factors of these groups (see Remark \ref{remark:group-effect-sparsity} and Section S3 of the Supplementary Material).
Some intuition can be gained by considering the extreme case of one active group, as the GOA endows one with the opportunity to study the factors of this group with the same dataset since the effect aliasing among them is small due to the high strength of the design projected onto this group.
The analysis strategy is different from that for traditional group screening, which typically requires two sets of experimentation, one for identifying active groups and the other for studying the factors of the active groups.

Another application where GOAs could play a role is the experiments for dimensional analysis \citep{Albrecht-etal-2013,Shen-etal-2014}.
In dimensional analysis, factors are divided into groups such that those in the same group can be formulated as a dimensionless factor.
By arranging the factors such that each dimensionless factor corresponds to a group of the GOA, more levels of the dimensionless factors can be observed, which would be advantageous for studying the effects of the dimensionless factors.

\subsection{Possible directions for future work}
For applications where more levels are needed, one may expand the levels of a GOA in the same way as constructing orthogonal array-based Latin hypercubes \citep{Tang-1993}.
Alternatively, one can also apply the technique of \cite{Sun-Tang-2017} to generate column-orthogonal designs from some GOAs presented in this paper.
We illustrate this with the GOA$(32, 8 \times 3, 3 \times 3, 2, 2)$ presented in Table S1.
Denote the centered version of this design (i.e., the two levels are denoted by $-1/2$ and $+1/2$) by $D=(D_1, D_2, D_3)$ and write $D_i = (D_{i1}, D_{i2})$ such that $D_{i1}$ and $D_{i2}$ each have 4 columns for $i=1,2,3$.
Let
$$
Q = \begin{bmatrix}
4	&	-2	&	-1	&	\phantom{-}0	\\
2	&	\phantom{-}4		&	\phantom{-}0		&	\phantom{-}1	\\
1	&	\phantom{-}0		&	\phantom{-}4		&	-2	\\
0	&	-1	&	\phantom{-}2		&	\phantom{-}4
\end{bmatrix}
\quad \mbox{and}	\quad
D_{ij}' = D_{ij} Q
$$
for $i=1,2,3$ and $j=1,2$.
Let $D_i' = (D_{i1}', D_{i2}')$ for $i=1,2,3$.
Then $D'=(D_1', D_2', D_3')$ is a column-orthogonal design of 32 runs for 24 factors with 8 levels.
In addition, since $D_i$ has strength 3, it can easily be proved that the elementwise product of any two columns in $D_i'$ is orthogonal to another column in $D_i'$ for $i=1,2,3$.
On the other hand, it is known that strong orthogonal arrays \citep{He-Tang-2013} are more space-filling than ordinary orthogonal array-based designs.
It is therefore interesting to explore how to construct strong orthogonal arrays by using GOAs as base arrays.

Many GOAs studied in this paper have constant group sizes.
In practice, we may also encounter situations where the groups have different numbers of factors.
One can drop columns from a GOA$(N, m \times g, t \times g, s, t_0)$ to obtain variable group sizes.
For example, all regular minimum aberration OA$(81, m, 3, t)$'s given in \cite{Xu-2005} for $m \leq 10$ are embedded in a regular OA$(81, 11, 3, 2)$.
Hence, using the GOA$(81, 11 \times 3, 2 \times 3, 3, 2)$ in Table S5, we can construct a GOA$(81, (m_1, m_2, m_3), (t_1, t_2, t_3), 3, 2)$ such that every group has minimum aberration among all regular designs for any choice of $m_i \leq 11$ ($i=1,2,3$).
Despite being simple and useful, this approach of dropping columns obviously cannot cover all scenarios.
It is thus a potential direction to construct GOAs with variable group sizes for future research.


Another important topic is the use of GOAs in analysis of computer experiments and beyond.
Besides the universal kriging model used in Section \ref{sec:simulation}, another approach to analyze the data, as mentioned in Section \ref{sec:introduction}, is to use a Gaussian process with a blocked additive kernel.
In such a model, the Gaussian process can be seen as a sum of several independent Gaussian processes where each process is defined over several factors called a block.
It is intuitive to arrange the factors in such a way that those corresponding to the same block enjoy better space-filling properties in the experimental design.
An interesting research problem is to examine how GOAs are connected to blocked additive kernels, as done in \cite{Lin-Morrill-2014}, which showed the advantages of designs of variable resolution in model selection of linear models.

\section*{Supplementary Materials}
Supplementary material available online includes all the proofs of theoretical results, design tables, and other technical details of the paper.
\par
\section*{Acknowledgements}
The authors would like to thank an Editor, an AE and two referees for their helpful comments which greatly improved the paper.
Chen is supported by National Natural Science Foundation of China, Grant No.~12401325.
He was supported by National Natural Science Foundation of China, Grant No.~11701033.
Lin was supported by Discovery Grant from the Natural Sciences and Engineering Research Council of Canada.
Sun is supported by National Natural Science Foundation of China (Nos.~12371259 and 11971098) and National Key Research and Development Program of China (Nos.~2020YFA0714102 and 2022YFA1003701). Fasheng Sun is the corresponding author.
\par


\bibhang=1.7pc
\bibsep=2pt
\fontsize{9}{14pt plus.8pt minus .6pt}\selectfont
\renewcommand\bibname{\large \bf References}
\expandafter\ifx\csname
natexlab\endcsname\relax\def\natexlab#1{#1}\fi
\expandafter\ifx\csname url\endcsname\relax
  \def\url#1{\texttt{#1}}\fi
\expandafter\ifx\csname urlprefix\endcsname\relax\def\urlprefix{URL}\fi

\bibliographystyle{chicago}      
\bibliography{paper-ref}   

%
%
%
%
%

\vskip .65cm
\noindent
Guanzhou Chen \\
School of Statistics and Data Science, LPMC \& KLMDASR, Nankai University
\vskip 2pt
\noindent
E-mail: gzchen@nankai.edu.cn
\vskip 2pt

\noindent
Yuanzhen He \\
School of Statistics, Beijing Normal University
\vskip 2pt
\noindent
E-mail: heyuanzhen@bnu.edu.cn
\vskip 2pt

\noindent
C.~Devon Lin \\
Department of Mathematics and Statistics,  Queen's University
\vskip 2pt
\noindent
E-mail: devon.lin@queensu.ca
\vskip 2pt

\noindent
Fasheng Sun \\
KLAS and School of Mathematics and Statistics,  Northeast Normal University
\vskip 2pt
\noindent
E-mail: sunfs359@nenu.edu.cn

\end{document}